# Learning the Shoreline: A Very High-Resolution Approach to Reef Island Dynamics.

T. Fischer[1], B. Stoll[1]

[1] T. Fischer, UMR 241 SECOPOL, University of French Polynesia, Tahiti, French Polynesia – tobias.fischer@doctorant.upf.pf

[2] B. Stoll, UMR 242 SECOPOL, University of French Polynesia

**Abstract.** Pacific atoll islets are often described as stable in global-scale studies, typically based on long-term shoreline proxies such as vegetation line or morphometrics like planform surface area. While informative, these approaches can obscure short-term, localized coastal dynamics – including changes in island shape and position – that are critical for ecosystem function, cultural practices, and coastal infrastructure resilience.
This study presents a transferable, automated approach to shoreline monitoring using very high-resolution Pléiades imagery and a XGBoost classifier. The method integrates spectral indices and textural features to delineate the outer limit of emerged land, including vegetated areas, beaches, man-made surfaces, and beach rock. This shoreline definition supports fine-scale, spatially explicit monitoring of reef island dynamics, even in morphologically complex environments. Developed and tested on multiple atolls in French Polynesia (Tetiaroa, Tikehau, Hao, and Puka Puka), the model achieves high accuracy (mean Intersection over Union ≈ 0.99; Mean Absolute Positional Error ≈ 1.28 m) and demonstrates strong performance on both training and held-out sites, validating its spatial transferability. The extracted shorelines reveal subtle but significant island-scale changes in extent, configuration, and spatial position that remain undetected by conventional shoreline proxies and surface metrics. By enabling high-precision, scalable shoreline monitoring, this method provides a more nuanced understanding of atoll change processes. It supports Pacific efforts to move beyond narratives of passive loss toward frameworks of resilience and adaptation, while providing spatial tools tailored to low-lying island realities.

**Main Author Biography:** Tobias Fischer is a PhD candidate at the University of French Polynesia (UMR 241 SECOPOL). He holds degrees in International Area Studies (MSc) and Business Administration (BSc), with training in geostatistics, remote sensing, and GIS. His research focuses on very high-resolution shoreline monitoring of Pacific atolls.

## 1 Introduction

Atoll islands are among the most exposed landforms to climate threats such as sea-level rise, coastal erosion, and extreme events. Their low elevation, narrow shape, and dynamic structure make them highly sensitive to both short-term disturbances and long-term shifts. Monitoring these changes is vital for evaluating habitability, infrastructure risks, and ecosystem health.

Classic assessments often rely on coarse imagery, planform metrics, or proxies like the vegetation line. While useful for long-term trends, they often miss localized shoreline changes – such as *motu* (Tahitian term for islet) reshaping, sand movement (erosion and accretion), or artificial expansion – that matter to local stakeholders.

Increased availability in very high-resolution (VHR) satellite imagery and advances in machine learning enable detailed shoreline monitoring. However, most studies remain site-specific, requiring local training data that limits scalability across the Pacific's diverse island systems [1].

This study evaluates a transferable shoreline classification approach using Pléiades imagery and an XGBoost model. Tested on diverse atoll settings in French Polynesia, it demonstrates high-precision, scalable monitoring. It is among the first to apply transfer learning to shoreline mapping on atolls, enabling cross-site classification without retraining.

## 2 Study Area and Data

### 2.1 Geographic Context and Atoll Selection

Four atolls in French Polynesia – Tetiaroa in the Society Islands, and Tikehau, Puka Puka, and Hao in the Tuamotu Archipelago – were chosen

for their contrasting geomorphologies, reef structures, and land use. These variations provide a robust test for the transferability of shoreline monitoring.

Climatic regimes vary slightly. The Tuamotus experience steady trade winds and low rainfall due to their flat, remote nature. In contrast, Tetiaroa receives higher and more variable rainfall due to its proximity to high volcanic islands in the Society group.

### 2.2 Atoll Characteristics

The four atolls also differ in their lagoon-ocean connectivity, shoreline configuration, and degree of human modification. These contrasts are important for evaluating model transferability.

Tetiaroa and Puka Puka are small closed atolls with no major passes. Tetiaroa is semi-managed, hosting an eco-resort and research station on one *motu*, while the others remain conserved. Puka Puka is minimally developed and largely covered by coconut plantations.

Tikehau and Hao are larger, open atolls, each with a major pass that enhances lagoon-ocean exchange. Tikehau includes both natural and built environments, including tourism infrastructure. Hao has undergone extensive military development, introducing artificial platforms and armoring that alter shoreline geometry.

### 2.3 Satellite Data

The image dataset includes Pléiades scenes acquired between 2016 and 2023. Four images were used for Tetiaroa (2016, 2019, 2022, and 2023), while a single image from 2022 was used for each of the three other atolls: Hao, Puka Puka, and Tikehau. All images were captured between May and July to minimize seasonal effects and delivered as Level 2A products (atmospherically corrected, co-registered, and pan-sharpened).

Scenes with minimal cloud and haze were prioritized. Partial cloud cover affected parts of Tetiaroa (2016, 2023) and Tikehau (2022), but most shorelines were cloud-free. For Hao and Tikehau, analysis focused on representative sections near main villages and adjacent *motu* to balance coverage and computational cost.

## 3 Methods

### 3.1 Shoreline Reference Preparation

The term shoreline is conventionally defined as the physical interface between land and sea [2]. In remote sensing applications, this often corresponds to the instantaneous land–water boundary visible in an image [3]. However, this interface is inherently dynamic and scale-dependent, influenced by seasonality, tidal stage, wave run-up, sediment movement, and geomorphic variability. On atolls, additional complexity arises from narrow *motu*, irregularly shaped, water-retaining conglomerate platforms, and the presence of vegetation shadows – all of which can obscure the true extent of emerged land in VHR imagery.

The shoreline is operationally defined as the outer edge of emerged land visible in Pléiades imagery, encompassing vegetated zones, bare beaches, reef platforms, and artificial structures, but excluding submerged reef flats. Depending on land cover, the mapped shoreline corresponds to the beach toe, the vegetation line, or the stability line.

### 3.2 Classification Approach

A set of spectral indices and texture attributes was derived from the pan-sharpened Pléiades imagery to support land–water classification. Common shoreline indices and GLCM texture features (mean, contrast, entropy) from the NIR band were computed using a 3×3 window [4, 5].

Training data were selected via a clustering-based strategy inspired by [6]. A representative region of interest was used to capture spectral variability, followed by separate k-means clustering for land and water classes. From each cluster, 1% of image pixels were randomly and equally sampled, yielding a balanced dataset split 80% for training and 20% for validation.

Model training was performed using the XGBoost classifier, with two-step Optuna-based hyperparameter tuning (initial coarse tuning,

followed by cross-validated refinement). The composite objective combined log-loss and macro-F1 score to balance accuracy and generalizability.

The final classification produced a binary land–water raster. Postprocessing removed small artifacts and filled internal gaps to ensure shoreline continuity. Shorelines were then extracted using a marching squares algorithm with linear interpolation and compared to ground truth references for accuracy assessment [5, 7].

### 3.3 Transfer Learning Setup

To evaluate the transferability of shoreline classification across space and time, a transfer learning strategy was applied. A single XGBoost model was trained on a composite dataset from Tetiaroa (2016, 2019, 2023), Hao (2022), and Puka Puka (2022). It was then tested on two held-out targets: Tetiaroa (2022), a temporally held-out scene, and Tikehau (2022), a spatially held-out atoll.

The same input features, sampling strategy, and shoreline extraction method were used across all configurations. Performance was evaluated using Intersection-over-Union (IoU), the proportion of shoreline predictions within 2 meters of the reference line (% within 2 m), and the 95th percentile of shoreline distance error. These metrics were used alongside conventional indicators such as MAPE and RMSE, which are more sensitive to extreme values. Percentile- and threshold-based indicators have also been employed in recent shoreline studies to complement standard accuracy metrics [5, 7, 8].

## 4 Results

The transfer learning model (TF) consistently matched or outperformed the per-image trained models (Original), including on temporally and spatially held-out scenes. The ‘Original’ models were trained and tested individually per image, whereas the TF model was trained once on five scenes and tested on two held-out targets. Table 1 summarizes shoreline classification accuracy across all test images. Fig. 1 illustrates this performance, showing *motu*-scale shoreline change on Tetiaroa.

The TF model maintained high Intersection-over-Union (IoU) scores—generally above 0.97—and exceeded 87% of predictions within 2 m of the reference shoreline in five of seven scenes. The 95th percentile error remained below 3.9 m in all but one case, confirming strong spatial delineation across varied atoll settings.

**Table 1.** Comparison of Original (O) vs. Transfer Learning (TF) model accuracy using IoU, %<2 m, and P95. Green = improvement, orange = deterioration, gray = no change. (* cloud corrected, + held-out scene)

| Atoll-Year | IoU (O) | IoU (TF) | %< 2m (O) | %< 2m (TF) | P95 (O) | P95 (TF) |
|---|---|---|---|---|---|---|
| Tetiaroa 2016* | 0.993 | 0.993 | 90.5 | 90.8 | 3.36 | 3.00 |
| Tetiaroa 2019 | 0.992 | 0.993 | 88.8 | 90.6 | 5.36 | 2.99 |
| Tetiaroa 2022+ | 0.994 | 0.992 | 93.2 | 88.4 | 2.28 | 2.87 |
| Tetiaroa 2023* | 0.994 | 0.994 | 93.9 | 93.5 | 2.15 | 2.20 |
| Puka Puka 2022 | 0.993 | 0.993 | 81.4 | 81.8 | 5.42 | 4.99 |
| Hao 2022 | 0.980 | 0.976 | 86.0 | 88.0 | 4.13 | 3.23 |
| Tikehau 2022*+ | 0.966 | 0.972 | 83.0 | 87.1 | 4.19 | 3.84 |

Tetiaroa 2019 showed a particularly strong transfer performance, with the TF model outperforming the Original across all metrics. The 95th percentile error fell from 5.36 m (O) to 2.99 m (TF), and the % within 2 m improved from 88.8% to 90.6%. Similar results were observed in Tetiaroa 2016 – a training image – demonstrating stable performance with notable improvements over the original model, particularly in the 95th percentile error and % within 2 m.

In contrast, Tetiaroa 2022, a temporally held-out scene, showed slightly reduced performance (e.g., P95 = 2.87 m vs. 2.28 m, % < 2 m = 88.4% vs. 93.2%), though all metrics remained within a high or acceptable range. The spatially held-out site, Tikehau 2022, showed strong generalization: the TF model outperformed the Original across all indicators, including a drop in P95 from 4.19 m to 3.84 m.

While Mean Absolute Positional Error (MAPE) and Root Mean Square Error (RMSE) are commonly reported in shoreline studies [5, 7, 8], they are sensitive to outliers and can overstate error in the presence of visual obstructions or classification ambiguity. In Tetiaroa 2016, for instance, MAPE dropped from 2.55 m to 1.28 m and RMSE from 8.96 m to 3.84 m after excluding cloud-impacted segments. Although computed for completeness, MAPE and RMSE were supplemented with more robust spatial indicators. Across all non-training test scenes, the TF model achieved a mean MAPE of 1.28 m and a mean RMSE of 3.42 m, ranging from 1.25 m (Tetiaroa 2023) to 8.07 m (Tikehau).

## 5 Discussion and Implications

Results confirm that transfer learning enables stable, accurate shoreline classification across diverse atoll settings and timeframes. The TF model consistently achieved high IoU scores (>0.97) and low MAPE values, averaging 1.28 m overall and 1.44 m for the two held-out scenes (Tetiaroa 2022 and Tikehau 2022), demonstrating strong spatial and temporal generalization across varying island morphologies.

This consistency is especially valuable for atolls, where shorelines are narrow, complex, and sensitive to both natural and human pressures. The model's performance without local retraining aligns with previous findings [5], which show that pretrained classifiers reduce workload, enhance repeatability, and enable scalable shoreline mapping. These advantages are critical in Pacific island contexts, where annotated training data is limited and the need for adaptable, low-maintenance models is high, despite the growing demand for local environmental monitoring.

VHR imagery proved essential for capturing the fine-scale geomorphic variability of reef islands. As noted by [9], VHR imagery allows accurate delineation of narrow landforms and captures shoreline changes that coarser data miss. In this analysis, Pléiades images revealed subtle shoreline shifts, sediment redistribution, and *motu* reshaping – even when total land area remained unchanged. These patterns align with long-term shoreline dynamics observed on Tetiaroa (1955-2023) [10]. For example, on the *motu* Tahuna Iti in Tetiaroa, extracted shorelines revealed localized erosion and accretion patterns not reflected in global-scale area metrics (see Fig. 1), highlighting the limits of planform surface metrics alone and the value of direct shoreline mapping.

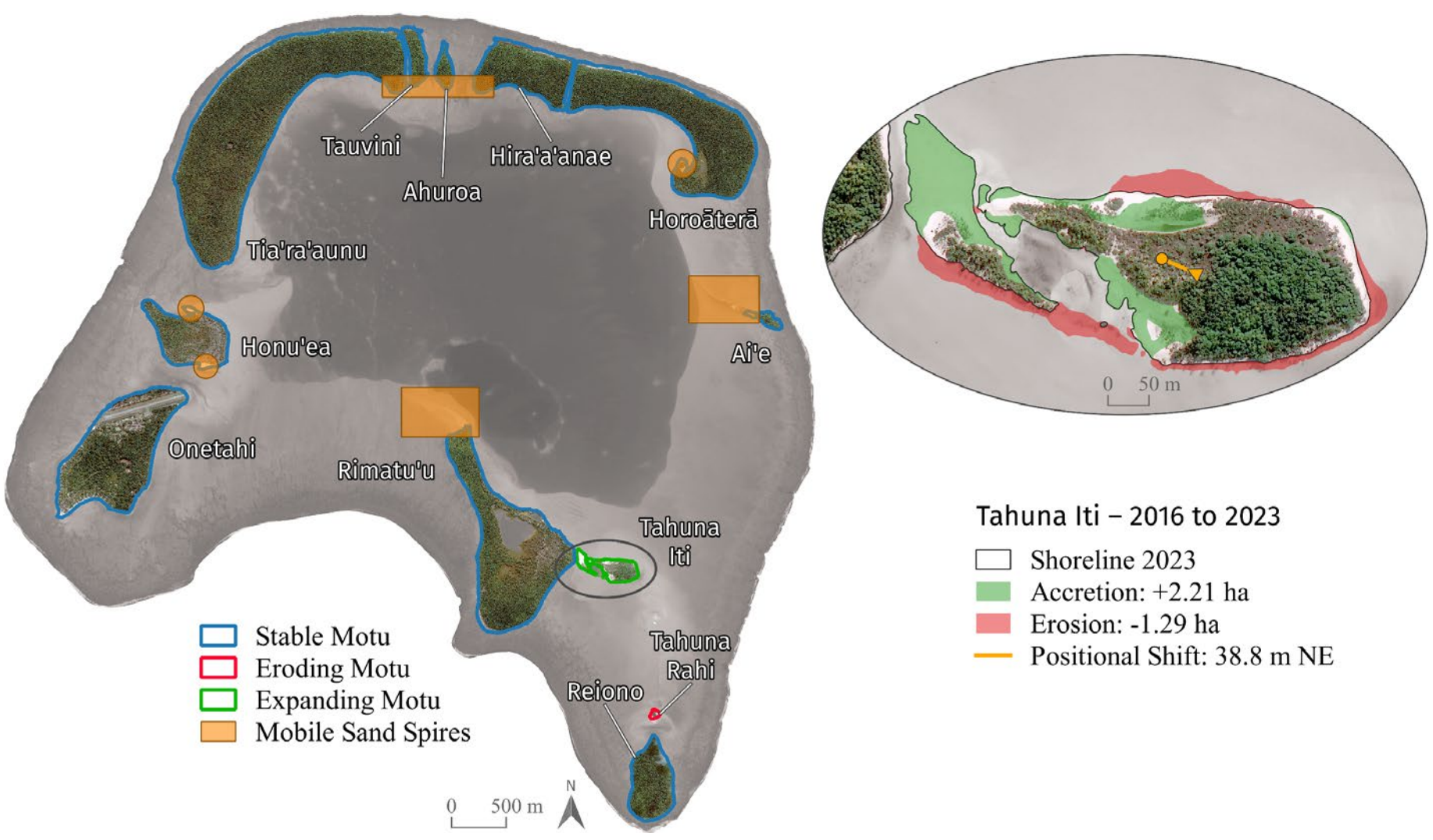

**Fig. 1.** *Motu*-scale shoreline changes on Tetiaroa between 2016 and 2023. Main map shows status of each *motu*: expanding (green), eroding (red), or stable (blue), with mobile sand spires in orange. The inset highlights Tahuna Iti, where 2.21 ha of accretion and 1.29 ha of erosion occurred, along with a 38.8 m northeastward shift of the shoreline over the 7-year period.

While robust across sites, some limitations remain. Cloud and shadow artifacts required manual correction, especially where spectral mixing obscured the shoreline. Corrections for water-level variability were not applied due to the absence of tide gauge data and beach profiles, which are often unavailable in remote atoll settings. These inputs are essential for implementing tidal normalization, as demonstrated in studies incorporating tide-based shoreline corrections [5]. Some ambiguity also persists in transitional zones, such as tide pools and overhanging vegetation. Future work should broaden the training dataset to improve model adaptability across island types and timeframes.

The method provides an operational workflow aligned with regional adaptation goals. It supports moving beyond binary loss/gain narratives toward spatially nuanced insights – enabling resilience strategies grounded in fine-grained observation.

## 6 Conclusion

This study demonstrated the potential of transfer learning and VHR satellite imagery for automated, accurate shoreline detection in atoll environments. The proposed approach maintained high spatial accuracy, with a mean positional error of ≈ 1.28 m, and achieved robust accuracy on held-out sites without local data or retraining, demonstrating field-ready generalization. By integrating spectral and textural information, the method captures subtle shoreline changes that remain invisible in conventional metrics like land area. Despite some limitations related to cloud cover, tidal variability, and shoreline ambiguity, the workflow supports scalable and spatially explicit monitoring suited to the realities faced by Pacific islands. This workflow enables actionable, fine-scale monitoring that meets the operational needs of Pacific islands' managers.

## Acknowledgements

This research was conducted as part of a PhD project funded by the French government and hosted by UMR 241 SECOPOL at the University of French Polynesia. The authors gratefully acknowledge the logistical, technical, and financial support provided by Tetiaroa Society, the research units GePaSud and Espace-Dev New Caledonia, and the Topography Section of the French Polynesian Land Affairs Department. Special thanks are extended to the Agence Universitaire de la Francophonie for supporting participation in the 6th PIURN Conference.